\documentstyle[12pt]{article}

\newfont{\Mb}{msbm10}

\newcommand{\dx}{{\rm d}x}
\newcommand{\dy}{{\rm d}y}

\newcommand{\Pd}[2]{{{\partial #1}\over{\partial #2}}}

\begin{document}
\setcounter{equation}{0}
\setcounter{figure}{0}
\setcounter{table}{0}

\hspace\parindent
\thispagestyle{empty}

\bigskip
\bigskip
\bigskip
\begin{center}
{\LARGE \bf Improving a family of Darboux methods}
\end{center}
\begin{center}
{\LARGE \bf for rational second order ordinary}
\end{center}
\begin{center}
{\LARGE \bf differential equations}
\end{center}

\bigskip

\begin{center}
{\large L.G.S. Duarte and L.A.C.P. da Mota \footnote{E-mails:
lduarte@dft.if.uerj.br and damota@dft.if.uerj.br} }
\end{center}

\bigskip
\bigskip
\centerline{\it Universidade do Estado do Rio de Janeiro,}
\centerline{\it Instituto de F\'{\i}sica, Depto. de F\'{\i}sica Te\'orica,}
\centerline{\it 20559-900 Rio de Janeiro -- RJ, Brazil}

\bigskip
\bigskip

\bigskip
\bigskip

\abstract{We have been working in many aspects of the problem of
analyzing, understanding and solving ordinary differential
equations (first and second order). As we have extensively
mentioned, while working in the Darboux type methods, the most
costly step of our methods and algorithms of solution is the determination of
Darboux polynomials for the associated differential operators.
Here, we are going to present some algorithms to greatly reduce the time expenditure in determining these needed Darboux polynomials.  Some of them are based on  a detailed analysis of the general
structure of second order differential equations regarding the
associated differential invariants. In order to perform this analysis, we produce a theorem
concerning the general form for the differential invariants in terms of the Darboux polynomials.}

\bigskip
\bigskip
\bigskip
\bigskip
\bigskip
\bigskip



\newpage

%%%%%%%%%%%%%%%%%%%%%%%%%%%%%%%%%%%%%%%%%%%%%%%%%%%%%%%%%%%%%%%%%%%%%%%%%%%%%%%%
\section{Introduction}
\label{intro}

The differential equations (DEs) are the most widespread way to
formulate the evolution of any given system in many scientific
areas. Therefore, for the last three centuries, much effort has
been made in trying to solve them.

In the first approaches to solving DEs, the methods consisted on,
for a given DE or class of them, trying to use a specific method
to deal with that particular case. With the cumulation of such
experiences, a classificatory system was produced where a specific
set of rules was designed for  certain classes of DEs. By the mid
XIX$^{th}$ century, this situation had established itself as the
way to deal with DEs.

Latter on, more comprehensive methods (non-classificatory) have
been developed. Just to mention a few, in the second half of the
XIX$^{th}$ century S. Lie
\cite{step,bluman,olver,Olver2,Ibragimov}, by the development of
his continuous groups of symmetries, managed to unify this whole
lot of particular approaches to deal with ordinary DEs (ODEs) into
a single theoretical framework. Another type of approach we would
like to highlight is the Darboux type approach \cite{PS,Shtokhamer,collins,chris1,chris2,llibre}.

In the Darboux type approach front, it is worth mention that a
(semi) algorithmic approach applicable to solving first order
ordinary differential equations was made by M. Prelle and M.
Singer \cite{PS}. The attractiveness of the Prelle-Singer (PS)
method lies in that, if the given first order ODE has a solution
in terms of elementary functions, the method guarantees that this
solution will be found (though, in principle it can admittedly
take an infinite amount of time to do so). The original PS method
was built around a system of two autonomous first order ODEs of
the form $\dot{x} = N(x,y)$, $\dot{y}=M(x,y)$ with $M$ and $N$
polynomials in ${\it C}[x,y]$ or, equivalently, a rational first
order ODE of the form\footnote{
From now on, we will use that $dy/dx = z$.} $z\equiv dy/dx = M(x,y)/N(x,y)$.

We have been working on analyzing and solving systems of first and
second order differential equations (1ODEs and 2ODEs,
respectively) from a numerical point of view \cite{ndyn}, using
Lie methods \cite{nossolie1,nossolie2} and Darboux type approaches
\cite{PS2,firsTHEOps1,secondTHEOps1,nossoPS1CPC,iccmse2005,JCAM,AMC,JMP,3dpol}. For
this latter class of methods, we have been developing (semi)
algorithms to deal with classes of ODEs. In these algorithms, one
fact has been always present: the most (computationally) costly
step is the determination of the associated Darboux polynomials.

Based on that realization, here we will be focused on speeding the
process of finding Darboux polynomials for a class of ODEs of our
interest. In particular, in this paper, we will talk about a class
of rational 2ODEs.

This finding of the Darboux polynomials (quicker than before, via
other methods) is useful on its own right. But, in particular, its
very compelling to the class of Darboux type methods we have been
working with. To exemplify this, we will use the algorithm
developed by us in \cite{JMP}.

The paper can be summarized as follows: In the next section, we
will briefly introduce the main features of the method we will use
as model for the methods benefitted from our here presented method
to speed the finding of the Darboux polynomials. In section
(\ref{NxyMz}), we will present a part of the method where one can
find, in a very straightforward manner, the Darboux polynomials
for a class of 2ODEs by inspection. Next, in section (\ref{N-Inv}), we will
present the part where a deeper analysis of the structure of the
2ODE is needed to extract the Darboux polynomials. In each of the
above mentioned two sections, we will present examples of the
application of our proposed method. Finally, we will present our
conclusions.

\section{Summary of the Darboux type approach we will use as model for a the benefited approaches}
\label{introNEW}

In \cite{PS2} we developed an extension of the Prelle-Singer
method \cite{PS} and, in that paper we proposed to use an unknown
function (that we called $S$) in order to make the
1-form\footnote{This 1-form is associated with the rational 2ODE
$z'=\phi(x,y,z)$, where $\phi$ is a rational functions of
$(x,y,z)$.} $\,\phi(x,y,z)\,dx-dz\,$ proportional to an exact
1-form. In \cite{AMC} we constructed a semi-algorithm to determine
the $S$-function for an 2ODE presenting an elementary first
integral. In \cite{JMP}, we built on a new, more efficient
algorithm, and established a set of theoretical results that makes
the basis of the algorithm. The work here presented will, in the
majority, be concerned in improving one step of the algorithm:
namely the finding of Darboux polynomials (that is an essential
part of the procedure).

In \cite{JMP}, we use a variation of the idea of the function $S$.
Basically, we use (in order to span the space of null 1-forms)
that:

If the 2ODE is given by:
\begin{equation}
\label{2ode} \frac{d^2y}{dx^2} = \phi (x,y,z) =
\frac{M(x,y,z)}{N(x,y,z)},
\end{equation}
where $M$ and $N$ are polynomial functions of $(x,y,z)$.

We then construct a differential operator $D$,
\begin{equation}
\label{Doperator}
 D \equiv N\,\partial_{x} + z\,N \partial_{y} + M\,
\partial_{z},
\end{equation}
extracted from the second order ODE, and their corresponding
Darboux polynomials and co-factors will be the building blocks of
the integrating factor and, ultimately, of the differential
invariant for the 2ODE.

We can write (\ref{2ode}) in the form
\begin{equation}
\label{sys_charact2} \frac{dx}{N} = \frac{dz}{M}
\end{equation}
and since $z\,dx = dy$, we have that the 1-forms $\alpha$ and
$\beta$ defined by $\alpha \equiv M\,dx - N\,dz$ and $\beta
\equiv z\,dx - \,dy$ are null over the solutions of the ODE
(\ref{2ode}), i.e., over the solutions,
\begin{equation}
\label{aeb} \alpha \equiv M\,dx - N\,dz = 0\,,\,\,\, \beta \equiv
z\,dx - \,dy = 0.
\end{equation}
\noindent From these results we have the following:

\bigskip

\noindent {\it If $I(x,y,z)$ is a first integral of the ODE
$($\ref{2ode}$)$ then the 1-form $dI$ is a vector in the subspace
spanned by the 1-forms $\alpha$ and $\beta$ defined above, i.e.,
\begin{equation}
\label{linear3} dI = r\, \alpha + s\, \beta
\end{equation}
\noindent where $r$ and $s$ are functions of $(x,y,z)$.}

\bigskip

From (\ref{linear3}) we have
\begin{equation}
\label{linear4} I_x \, dx + I_y \, dy + I_{z} \, dz = r \,
(M\,dx - N\,dz) + s \, (z\,dx - \,dy)
\end{equation}
\noindent implying that
\begin{eqnarray}
\label{IxIyIz}
I_x & = & r\,M + s\,z \nonumber \\
I_y & = & -\,s \nonumber \\
I_{z} & = & -\,r\,N
\end{eqnarray}
\noindent Therefore, if we determine, $r$ and $s$ we can find $I$
via quadratures using the equation:
\begin{eqnarray}
\label{IntI}
I(x,y,z) = \int I_x\,\dx+\int\!\left[ I_y - \Pd{}{y}\int I_x \dx\right]\dy + \nonumber \\
\int\!\left( I_{z} - \Pd{}{z}\left[ \int I_x\,\dx+\int\!\left[
I_y - \Pd{}{y}\int I_x \dx\right]\dy   \right] \right) dz
\end{eqnarray}
%from (\ref{IxIyIz}), we get:
%\begin{eqnarray} \label{intI} I(x,y,z) = \int \!\left(r\,M+s{\it
%-\int\!\left[s+\Pd{}{y}\int \!\left(r\,M+sz\right)\dx\right]\dy- \nonumber \\
%\int\!\left\{r\,N+\Pd{}{z}\left[\int\!(r\,M+sz)\dx
%-\int\!\left[s+\Pd{}{y}\int\!(r\,M+sz)\dx\right]\dy\right)\right\}\dz.
%\end{eqnarray}

\bigskip

In \cite{JMP}, in order to generate an operational semi-algorithm,
i.e. being able to determine $I_x,I_y$ and $I_z$, we have used the
class of ODEs where $s/r$ $(r$ and $s$ defined by equation
$($\ref{linear3}$))$ is a rational function of $(x,y,z)$ $($
i.e., $s/r = P/Q$ where $P$ and $Q$ are polynomials that do not
have any common factors$)$.

 Also in \cite{JMP}, we have demonstrated many results, including two theorems, that allow us to
construct a semi-algorithm to find elementary first integrals of a
class of rational second order ODEs (via a Darboux-type
procedure), thus generalizing the correspondent method developed
by Prelle and Singer for first order ODEs. Brutally summarizing
the important results for us here, we have found that:

\bigskip
We can write:
\begin{equation}
R \equiv r/Q = \prod_i v^{m_i}_i \Rightarrow \frac{D[R]}{R} = \sum
m_i\,\frac{D[v_i]}{v_i} = \sum m_i\,g_i
\end{equation}
\noindent where $R$ is the integrating factor and $v_i$ are
irreducible Darboux polynomials (in $(x,y,z)$) of the $D$
operator and the $g_i$ are the corresponding co-factors. Then,
from the compatibility conditions, $(I_{xy}=I_{yx},
I_{xz}=I_{zx}$ and $I_{yz}=I_{zy} )$, we can write:
\begin{equation}
\label{PDr2new} P\,\left(\frac{D[R]}{R}\right) = P\,\sum m_i\,g_i
= -D[P] - Q\,(N\,M_y-M\,N_y).
\end{equation}
\noindent and
\begin{equation}
\label{QDr2new} Q\,\left(\frac{D[R]}{R}\right) = Q\,\sum m_i\,g_i
= -D[Q] - P - Q\,(N_x+N_y\,z+M_{z})
\end{equation}

Ultimately, these ideas led to:

\begin{equation}
\label{eqdI3} dI = R\,\left[(M\,Q + z\,P)\,dx + (-\,P)\,dy +
(-\,N\,Q)\,dz\right]\,,
\end{equation}
and
\begin{eqnarray}
\label{IxIyIz2}
I_x & = & R\,(M\,Q + z\,P) \nonumber \\
I_y & = & -\,R\,P \nonumber \\
I_{z} & = & -\,R\,N\,Q
\end{eqnarray}
\noindent

In \cite{JMP}, equations (\ref{PDr2new},\ref{QDr2new}) were the
basis of our procedure. So, by determining the Darboux polynomials
and solving them, we would have found $P,Q$ and $R$, thus enabling
us to use equations (\ref{IntI}, \ref{IxIyIz2}) and find the
invariant.

The great advantage of our method is that it converts the search
for first integrals into solving (essentially) first degree
algebraic equations (in the same way as Prelle and Singer did for
first order ODEs). As a consequence of this, our approach is
capable of analyzing the integrability regions for the seconde
order ODE (for the case where it presents undetermined
parameters). To do this we have to `add' some (or all) parameters
as variables into the algebraic system we have to solve. The
solutions exists for values where there is an integration possible
\footnote{In \cite{iccmse2005,AMC} we had already introduced this procedure}.
Furthermore, the procedure is semi-algorithmic and, given enough
time, if the solution exists, it will find it (as does the
Prelle-Singer approach for first order ODEs).

It is worth to point out that another positive aspect (that will
become clear just a few steps ahead, on the next section) is that
we use (in our method presented in \cite{JMP} the differential
operator given by equation (\ref{Doperator}), instead of the
(probably) more commonly used one:
\begin{equation}
\label{Doperator_old}
 D \equiv \partial_{x} + z\,\partial_{y} + \phi\,
\partial_{z},
\end{equation}
where $\phi$ has the same meaning as in equation ({\ref{2ode}).

The Darboux polynomials for the operator given by equation
(\ref{Doperator}) are used (as explained above) in the algorithm (together with their
associated co-factor) in order for us to find the differential
invariants for the 2ODE in question. Next, let us try to produce ways to find them more quickly.

% {\bf \Large Just to situate the readers, avoiding boring them to death, let us summarize the main results}

\section{Darboux polynomials as straightforward factors in the numerator and-or in the denominator}
\label{NxyMz}

In this section, we will extract information regarding the Darboux
polynomials, correspondent to the operator (\ref{Doperator})
related to the 2ODE being studied in a very direct way. This,
although a simple procedure, will prove essential to solve (or at
least reduce) some ODEs.

As mentioned in the previous section, the fact that our method
\cite{JMP} uses the differential operator given by equation (
\ref{Doperator} ) is very advantageous. One can see, just by
inspecting this operator, that if the denominator of the 2ODE (\ref{2ode}),
 i.e. $N$, is a function of $(x,y)$ only, it will be
certainly a Darboux polynomial of the D-operator defined in
(\ref{Doperator}). Actually, for that matter, any factor of $N$
that depends only on $(x,y)$ will do. Analogously, for the case of the
numerator in (\ref{2ode}) but, this time, regarding functions of $(z)$
only. The numerator or any factor of it (that is a function of $z$
alone) is a Darboux polynomial of the D-operator
(\ref{Doperator}).

In the next two sub-sections, we will study each of these cases.

\subsection{Darboux polynomials as functions of $(x,y)$ only}
\label{Nxy}

In order to analyze this case, let us first re-write the
D-operator (eq. \ref{Doperator}):

\begin{eqnarray}
 D &\equiv& N\partial_{x} + z\,N \partial_{y} + M\,
\partial_{z}, \nonumber
\end{eqnarray}

As explained above, the case of interest here is the one where $N$
is a function only of $(x,y)$ or it has factors that are so. To
cover all the cases, consider the following general form for $N$:

\begin{equation}
N =  (\prod_{i=1}^k n_i(x,y) ) {\cal N}(x,y,z)
\end{equation}
where $n_i(x,y)$ and ${\cal N}(x,y,z)$ are polynomials in $(x,y)$
and $(x,y,z)$ respectively. It is easy to see that this general
form for $N$ covers all the cases of interest: namely, if the
there is no factor that is a function of $(x,y)$ one can consider
that the product $(\prod_i^k n_i(x,y) )=1$ and there is only
${\cal N}(x,y,z)$ left. This case will not be of interest here. On
the other hand, if the whole $N$ is a function of $(x,y)$ only,
this can be translated by saying that the product $(\prod_i^k
n_i(x,y) )=n_1(x,y)$ and ${\cal N}(x,y,z)=1$. Of course, the case
where there are a few factors is covered with $k \ge 1$.

These cases are listed bellow with the general expressions for the
Darboux polynomials and the associated co-factors:

\bigskip
\begin{itemize}
\item {\it {\bf case 1:} where the whole $N$ is a polynomial in $(x,y)$}
\begin{equation}
N=n_1(x,y) \rightarrow D[n_1]= n_1\,(\frac{\partial
n_1}{dx})+n_1\,z\,(\frac{\partial n_1 }{dy})
\end{equation}
\noindent
So, if $v_1=n_1$ is a Darboux polynomial of $D$, the
associated co-factor is:
\begin{equation}
g_1=(\frac{\partial n_1}{dx})+z\,(\frac{\partial n_1 }{dy})
\end{equation}

\bigskip
\item {\it {\bf case 2:} where there are one or more factors in $N$ that are polynomials without $z$}

\begin{eqnarray}
N &=& \left(\prod_{i=1}^k n_i(x,y) \right) {\cal N}(x,y,z)
\rightarrow  \nonumber \\
D[n_j] = &=&\left(\prod_{i=1}^k n_i(x,y)
\right) {\cal N}(x,y,z)\, \left((\frac{\partial
n_j}{dx})+z\,(\frac{\partial n_j }{dy})\right)
\end{eqnarray}
\noindent where $1\le j \le k$. So, if we consider $v_1=n_j$ as a
Darboux polynomial of $D$, the associated co-factor is:
\begin{equation}
g_1=\left(\prod_{i=1,i\ne j}^k n_i(x,y) \right) {\cal N}(x,y,z)\,
\left((\frac{\partial n_j}{dx})+z\,(\frac{\partial n_j
}{dy})\right)
\end{equation}
\end{itemize}

\subsection{Darboux polynomials as functions of $(z)$ only}
\label{Mz}

This situation is analogous to the one just dealt with. So we will
begin the exposition in a similar fashion,

Again, let us reffer to the D-operator given by eq.
\ref{Doperator}.

The focus now is when $M$ is a polynomial only on $(z)$ or it has
factors that are so. To cover all the cases, consider the
following general form for $M$:

\begin{equation}
M =  (\prod_{i=1}^k m_i(z) ) {\cal M}(x,y,z)
\end{equation}
where $m_i(z)$ and ${\cal M}(x,y,z)$ are polynomials in $(z)$ and
$(x,y,z)$ respectively. Again, one can see that this general form
for $M$ covers all the cases of interest: namely, if there is
no factor that is a function of $(z)$ one can consider that the
product $(\prod_i^k m_i(z) )=1$ and there is only ${\cal
M}(x,y,z)$ left. If $M$ is a polynomial on $(z)$ only, this can be
translated by saying that the product $(\prod_i^k m_i(z) )=m1(z)$
and ${\cal M}(x,y,z)=1$. Of course, the case where there is a few
factors is covered with $k \ge 1$.

These cases are listed bellow with the general expressions for the
Darboux polynomials and the associated co-factors:

\bigskip
\begin{itemize}
\item {\it {\bf case 1:} where the whole $M$ is a polynomial in $(z)$}
\begin{equation}
M=m_1(z) \rightarrow D[m_1]= m_1\,(\frac{\partial
m_1}{dz})
\end{equation}
\noindent So, if $v_1=m_1$ is a Darboux polynomial of $D$, the
associated co-factor is:
\begin{equation}
g_1=(\frac{\partial m_1}{dz})
\end{equation}

\bigskip
\item {\it {\bf case 2:} where there are one or more factors in $M$ that are polynomials without $(x,y)$}

\begin{equation}
M = \left(\prod_{i=1}^k m_i(z) \right) {\cal M}(x,y,z) \rightarrow
D[m_j] = \left(\prod_{i=1}^k m_i(z) \right) {\cal M}(x,y,z)\,
\left(\frac{\partial m_j}{dz}\right)
\end{equation}
\noindent where $1\le j \le k$. So, if we consider $v_1=m_j$ as a
Darboux polynomial of $D$, the associated co-factor is:
\begin{equation}
g_1=\left(\prod_{i=1,i\ne j}^k m_i(z) \right) {\cal M}(x,y,z)\,
\left(\frac{\partial m_j}{dz}\right)
\end{equation}
\end{itemize}

\subsection{Examples}
\label{examples1}

\subsubsection{first example}

Let us present an example that brings Darboux polynomials from $M$
(deppending on $(z)$) and $N$ (deppending on $(x,y)$):
\begin{equation}
\label{firstODE}
{\frac {d^{2}}{d{x}^{2}}}y =-1/2\,{\frac { \left(
2\,z+3 \right) \left( 3\,z{y}^{2}+z+x-{y}^{3}-y -1 \right)
}{x-{y}^{3}-y}}
\end{equation}
Using what we have been learning, we can see that we have two
Darboux polynomials: one from $M$ and another from $N$. Bellow we
will display both with their corresponding co-factors:

\begin{eqnarray}
v_1= 2\,z+3& \rightarrow &g_1=-2\,  \left(
3\,z{y}^{2}+z+x-{y}^{3}-y-1 \right) \nonumber \\
v_2=x-{y}^{3}-y& \rightarrow &g_2=2-6\,z{y}^{2}-2\,z
\end{eqnarray}
Using the method briefly described in section (\ref{introNEW}) one
conclude that, for this ODE, we have the following results for the
parameters and functions needed to find the differential invariant
for the ODE:

\begin{eqnarray}
\label{RPQ}
P&=& (3\, y^2+1)(3+2\,z)\nonumber \\
Q&=& 1 \nonumber \\
R&=& \frac{1}{(3+2\,z)(x-y^3-y)} .
\end{eqnarray}
and, using (\ref{IxIyIz2}) and (\ref{IntI}), we finally find:

\begin{equation}
\label{Inv}
 I = x-1+\ln  \left( {\frac {2\,z+3}{x-{y}^{3}-y}}
\right).
\end{equation}

It worth mention that the presence of the Darboux polynomial
$(x-{y}^{3}-y)$, with the cubic term, makes the regular process of
determinig it very ``expensive'' in time expenditure. After
applying the method here presented, which very quickly determined
the needed Darboux polynomials, the algorithm we introduced in
\cite{JMP} finds the results (\ref{RPQ}) and (\ref{Inv}) almost
instantly.

%second example

\subsubsection{second example}

This second example shows a case, where one of the Darboux needed
to determine, via the results presented on section \ref{introNEW},
is easily determined using the regular set of procedures (and this
one is not apparent on the format of the 2ODE in the fashion we
are advocating here) while the other, that would be very demanding
on time, is determined analyzing the denominator of the 2ODE. Let
us present the ODE\footnote{Incidently, althought it is not the
point here to compare every method of solving ODEs, this
particular ODE is not solved by the power methods and algorithms
implemented on the powerful Maple solve (release 10) }:

\begin{eqnarray}
\label{secondODE} \frac{d^2y}{dx^2} &=& \frac { \left(
2\,Cbxy+2\,Cb-xb \right){z}^{2}}{2\left( xy+1 \right)  \left(
Bax+Bbz+Bcy+b \right)} + \nonumber \\
&&\frac{\left(-yb+2\,Ca{x}^{2}y+2\,Cax+xcy+2\,c+2\,Cc{y}^{2}x+2\,Ccy-{x}^{2}a\right)z}{2\left(
xy+1 \right)  \left( Bax+Bbz+Bcy+b \right)} + \nonumber \\
&&\frac{yax+2\,a-{y}^{2}c}{ 2\left( xy+1 \right)  \left(
Bax+Bbz+Bcy+b \right) }
\end{eqnarray}
%\begin{equation}
%\label{secondODE} \tiny \frac{d^2y}{dx^2} = \frac { \left(
%2\,Cbxy+2\,Cb-xb \right)
%{z}^{2}+\left(-yb+2\,Ca{x}^{2}y+2\,Cax+xcy+2\,c+2\,Cc{y}^{2}x+2\,Ccy-{x}^{2}a\right)z+yax+2\,a-{y}^{2}c}{
%2\left( xy+1 \right)  \left( Bax+Bbz+Bcy+b \right) }
%\end{equation}

For this differential equation, if we we run the regular
procedures in order to find the Darboux polynomials, it is quickly
found one such polynomial.  But one needs more information than
that in order to be able to build an integrating factor for
(\ref{secondODE}). Using the analysis presented on section
(\ref{introNEW}) it is easy to see that $x\,y+1$ should be a
Darboux polynomial for the corresponding D-operator
(\ref{Doperator}). This combination of approaches generates the
following list of Darboux polynomials and co-factors:

\begin{eqnarray}
v_1= ax+cy+zb& \rightarrow &g_1=\left( 2\,zxcy+2\,yax+2\,a+2\,zc
\right) B+  \nonumber \\
& & \left( -2\,bzxy-2\,bz \right) C+xbz+yb \nonumber \\
 v_2=(x\,y+1)& \rightarrow &g_2=2\, \left( xy+1 \right)  \left(
zx+y \right)  \left( Bax+Bbz+Bcy+b
 \right)
\end{eqnarray}

Using this, one can determine $R$, $P$ and $Q$ corresponding to
(\ref{secondODE}) (see section (\ref{introNEW})):
\begin{eqnarray}
\label{RPQDOIS} P&=& \left(
2\,a{x}^{2}y+2\,ax+2\,bzxy+2\,bz+2\,c{y}^{2}x+2\,cy \right)
C-{x}^{2}a-xbz+ \nonumber \\
&&+xcy+2\,c \nonumber \\
Q&=& 1 \nonumber \\
R&=& -{\frac {1}{ \left( xy+1 \right)  \left( ax+bz+cy \right)}}.
\end{eqnarray}
and, using (\ref{IxIyIz2}) and (\ref{IntI}), we finally find:

\begin{equation}
\label{InvDOIS}
I = Bz+Cy+\ln  \left( {\frac {ax+bz+cy}{\sqrt {xy+1}}} \right).
\end{equation}

Analyzing the expression for $R$, we see that we needed two
Darboux polynomials in order to ``build'' it: namely $v_1=
ax+cy+zb$ and $v_2=(x\,y+1$ leading to $R=\sum_i v_i^{m_i}
\rightarrow  R = - \left( xy+1 \right)^{-1} \left( ax+bz+cy
\right)^{-1}$. We can also observe  that, even in this simple case
where one of the Darboux is of order one and the other of order
two, the solution of the ODE can elude many powerful techniques
(see footnote). The finding of the ``second'' Darboux polynomial
via the technique here suggested proves to be essential to render
the Darboux type approach presented in \cite{JMP} practical.

\section{Darboux polynomials from a deeper analysis of the 2ODE}
\label{N-Inv}

In this section, we will deepen our method through a more detailed
analysis of the structure of the differential equation we want to
solve (or, at least, reduce).

Let us do that beginning with the following:

\bigskip

Consider that

$$ I = I(x,y,z)$$
is a differential invariant corresponding to a certain 2ODE
$$\frac{d^2y}{dx^2} = \phi (x,y,z) = \frac{M(x,y,z)}{N(x,y,z)}$$

So, by using equation (\ref{Doperator}), one can write:

$$D[I]=0 \rightarrow  N\,\partial_{x}I + z\,N \partial_{y}I + M\,
\partial_{z}I = 0\rightarrow$$

\begin{equation}
\label{PHI}
\rightarrow \frac{M}{N} = \phi = -\frac{I_x+zI_y}{I_z}
\end{equation}

where $I_a = \partial_aI = \frac{\partial I(x,y,z)}{\partial a}$,
$a=x,y$ or $z$.

\bigskip
We may ask in what sense this is going to help our quest for
Darboux polynomials. Actually, we now have a great amount of
information regarding the general structure of the differential
invariant that, as we shall show now, will help us.

From \cite{PS}, we know that the differential invariant is of the form:
\begin{eqnarray}
\label{Inv_form_alg}
I &= &  W_0 + \ln
 \left( W_1  \right)
\end{eqnarray}
where $W_0$ and $W_1$ are algebraic functions of $(x,y,z)$.

Since we are interested in rational 2ODEs, one can easily see (from equations (\ref{PHI}) and (\ref{Inv_form_alg})) that we would have a relation between $W_0$ and $W_1$. So, from our experience in dealing with that question, we do not compromise the generality of the method much further if we use the differential invariant in the form:

\begin{eqnarray}
\label{Inv_form}
I &= &{\frac {A \left( x,y,z \right) }{{\cal D}
\left( x,y,z \right) }}+\ln
 \left( {\frac {B \left( x,y,z \right) }{C \left( x,y,z \right) }}
 \right)
\end{eqnarray}

where $A \left( x,y,z \right) ,{\cal D} \left( x,y,z \right) , B \left( x,y,z \right)$ and $C \left( x,y,z \right)$ are all polynomials.

Further still, as we will demonstrate below, if equation (\ref{Inv_form}) applies
, we have that ${\cal
D} \left( x,y,z \right) , B \left( x,y,z \right)$ and $C \left(
x,y,z \right)$ are Darboux polynomials of the D-operator
(\ref{Doperator}). This knowledge will prove essential in the
producing of new possibilities for determining the Darboux
polynomials for a given ODE.

%merda - começo da demonstração

\subsection{Darboux polynomials in the build-up of differential invariants}
\label{4ponto1}

Actually, in order to demonstrate the above mentioned results, let
us redefine a different differential invariant such that:

\begin{equation}
\label{Inv form 2}
I \rightarrow  {\cal I} = e^I = e^{\left( {\frac {A \left( x,y,z \right) }{{\cal D} \left( x,y,z \right)} }\right) } \left( {\frac {B \left( x,y,z \right) }{C \left( x,y,z \right) }} \right) \rightarrow {\cal I} = e^{\left( {\frac {A}{{\cal D}}} \right) } \left( {\frac {B}{C}} \right)
\end{equation}
where we have dropped the explicit notation of the $(x,y,z)$
dependence but it is still there.

Since ${\cal I}$ is an invariant, we have that:
\begin{equation}
\label{dinv}
{\frac{D[{\cal I}]}{{\cal I}}} = 0
\end{equation}

So, by using eq. ({\ref{Inv form 2}}), one gets:
\begin{eqnarray}
\label{dinv_expandido} {\frac{D\left[\frac{{\cal B}}{\cal C
}\right]   e^{\left( \frac {A}{{\cal D}} \right)} + \left(
\frac{{\cal B} }{{\cal C}}\right) D\left[\frac{{\cal A}}{\cal D
}\right] e^{\left( \frac {A}{{\cal D}} \right)} }{ \left(
\frac{\cal B}{\cal C} \right) e^{\left( \frac {A}{{\cal D}}
\right)} } } = 0
\end{eqnarray}
that finally results on
\begin{equation}
\label{dinv_expandido 2}
{\frac{D\left[\frac{{\cal B}}{\cal C
}\right]}{ \left( \frac{\cal B}{\cal C} \right) }} +
D\left[\frac{{\cal A}}{\cal D }\right] = 0
\end{equation}

Considering that $A, B, C, {\cal D}$ are all polynomials and that
we can therefore write them in terms of irreducible polynomials
$p_i$, we can use the following:

\begin{equation}
\label{dinv_expandido dois} \left( \frac{\cal B}{\cal C} \right) =
\prod_i p_i^{c_i}
\end{equation}
in turn, that leads to:
\begin{eqnarray}
\label{DBC}
\frac{D\left[\frac{{\cal B}}{\cal C }\right]}{ \left(
\frac{\cal B}{\cal C} \right) } &=&  {\frac{\sum_j c_j D[p_j] \left(\prod_{i \neq j} p_i^{c_i}\right) p_j^{c_j-1}}{\prod_kp_k^{c_k}}} \Rightarrow \nonumber \\
\Rightarrow \sum_j \left({\frac{c_jD[p_j]}{p_j}}\right) &=&
{\frac{\sum_j c_j D[p_j] \left(\prod_{k \neq j}
p_k\right)}{\prod_i p_i}}
\end{eqnarray}

Using equations (\ref{DBC}) and (\ref{dinv_expandido 2}), we get:

\begin{eqnarray}
\label{40}
\frac{\sum_j c_j D[p_j] \left(\prod_{k \neq j}p_k\right)}{\prod_i p_i} + D\left[\frac{{\cal A}}{{\cal D}}\right] = 0
\,\,\,\, \Rightarrow \nonumber \\
\sum_j c_j D[p_j] \left(\prod_{k \neq j} p_k\right) + \left( {\prod_i p_i} \right) D\left[\frac{{\cal A}}{{\cal D}}\right] = 0
\end{eqnarray}

Equation (\ref{40}) can finally be written as:

\begin{eqnarray}
\label{41} 
\sum_j c_j D[p_j] \left(\prod_{k \neq j} p_k\right)
&+& \left( \prod_i p_i \right) D\left[\frac{{\cal A}}{\cal D
}\right] = 0 \nonumber \\
\underbrace{\sum_j c_j D[p_j]  \left(\prod_{k \neq j} p_k\right)}
&+& \underbrace{\left( \prod_i p_i \right) D\left[\frac{{\cal
A}}{\cal D }\right]} = 0 \\
K_1 \,\,\,\,\,\,\,\,\,\,\,\,\,\,\,\,\,\,\,\,\,&+&
\,\,\,\,\,\,\,\,\,\,\,\,\,\,\,\,\,K_2 \nonumber
\end{eqnarray}

Since $K_1$ is a polynomial, so is $K_2$ $\Rightarrow$

\begin{eqnarray}
\label{42}
\Rightarrow K_2 = \left( \prod_i p_i \right)
D\left[\frac{{\cal A}}{\cal D }\right] & =&  \left( \prod_i p_i
\right) \left[\frac{ D[{\cal A}]{\cal D} - {\cal A} D[{\cal D}]
}{{\cal D}^2} \right] = \wp (x,y,z)
\end{eqnarray}
where $\wp (x,y,z)$ is a polynomial.

Now, remembering that $p_i$ are all irreducible, independent
polynomials, we have that:

\begin{equation}
\label{43}
\frac{\left( \prod_i p_i \right)}{{\cal D}^2}\neq
\verb"polynomial"
\end{equation}

That situation leads to two possibilities:

\subsubsection{$\left( \prod_i p_i \right)$ does not have common
factors with ${\cal D}^2$}
\label{4.1.1}

In this scenario, from equations (\ref{42} and \ref{43}), using
equation (\ref{dinv_expandido 2}) we have:

\begin{eqnarray}
\label{44}
\left[\frac{ D[{\cal A}]{\cal D} - {\cal A} D[{\cal D}]
}{{\cal D}^2} \right] &=& D\left[\frac{{\cal A}}{\cal D }\right]
\hbox{   is a polynomial} \nonumber \\
\Rightarrow {\frac{D\left[\frac{{\cal B}}{\cal C }\right]}{ \left(
\frac{\cal B}{\cal C} \right) }} &=& - D\left[\frac{{\cal
A}}{\cal D }\right] = \wp_1 (x,y,z)
\end{eqnarray}
where $\wp_1 (x,y,z)$ is a polynomial.

Let us analyze these results:

Consider the second (the one to the right) equality  of equation
(\ref{44}),
\begin{eqnarray}
\label{45}
D\left[\frac{{\cal A}}{\cal D }\right] &=& - \wp_1
(x,y,z) \Rightarrow \nonumber \\
\frac{ D[{\cal A}]}{{\cal D}} - \frac{{\cal A} D[{\cal D}]}{{\cal
D}^2} &=& - \wp_1 (x,y,z)\Rightarrow \nonumber \\
 - \frac{{\cal A} D[{\cal D}]}{{\cal
D}} &=&- \wp_1 (x,y,z)\,{\cal D} - D[{\cal A}]
\end{eqnarray}
Since the right-hand side of (\ref{45}) is obviously a polynomial
and, by construction, $\frac{{\cal A}}{{\cal D}}$ can not be
simplified (otherwise it would have been already), we may conclude
that:
\begin{eqnarray}
\label{DDarboux} \frac{D[{\cal D}]}{{\cal D}}& \hbox{is
polynomial}& \Rightarrow {\cal D}\hbox{ is a Darboux polynomial}
\end{eqnarray}
Actually, we can conclude more than that. If we write ${\cal D}$
as $\prod^{n_q}_jd_j^{k_j}$, where $d_j$ are irreducible
polynomials and $k_j$ integers, we have
\begin{equation}
\label{dqsq} \frac{D[{\cal D}]}{{\cal D}} =
\frac{D[\prod_{i=1}^{n_q} {\hbox{\scriptsize {\it
d}}}_i^{k_i}]}{\prod_{i=1}^{n_q} {\hbox{\scriptsize {\it
d}}}_i^{k_i}} = \sum_{i=1}^{n_q} k_i\, {\frac{D[{\hbox{\scriptsize
{\it d}}}_i]}{{\hbox{\scriptsize {\it d}}}_i}}.
\end{equation}
If we multiply (\ref{dqsq}) by ${\prod_{j=2}^{n_q}
{\hbox{\scriptsize {\it d}}}_j}\,$, we get
\begin{equation}
\label{dqsq1} \left({\prod_{j=2}^{n_q} {\hbox{\scriptsize {\it
d}}}_j}\right)\,\frac{D[{\cal D}]}{{\cal D}} =
k_1\,\left({\prod_{j=2}^{n_q} {\hbox{\scriptsize {\it
d}}}_j}\right)\,\frac{D[{\hbox{\scriptsize {\it
d}}}_1]}{{\hbox{\scriptsize {\it d}}}_1} + \sum_{i=2}^{n_q}
k_i\,\left({\prod_{j=2, j \neq i}^{n_q} {\hbox{\scriptsize {\it
d}}}_j}\right)\, D[{\hbox{\scriptsize {\it d}}}_i].
\end{equation}
Since he left hand side of (\ref{dqsq1}) and the second term on
the right hand side of (\ref{dqsq1}) are polynomials, we may
conclude that $k_1\,\left({\prod_{j=2}^{n_q} {\hbox{\scriptsize
{\it d}}}_j}\right)\,D[{\hbox{\scriptsize {\it
d}}}_1]/{\hbox{\scriptsize {\it d}}}_1$ is also a polynomial.
Considering that the ${\hbox{\scriptsize {\it d}}}$'s are
independent (by construction), the product ${\prod_{j=2}^{n_q}
{\hbox{\scriptsize {\it d}}}_j}$ can not cancel
${\hbox{\scriptsize {\it d}}}_1$. Therefore, we can conclude that
${\hbox{\scriptsize {\it d}}}_1|D[{\hbox{\scriptsize {\it
d}}}_1]$. In an analogous way, we have that ${\hbox{\scriptsize
{\it d}}}_i | D[{\hbox{\scriptsize {\it d}}}_i], i=2 \cdots n_q$.

So, what we can actually conclude is that ${\cal D}$ is formed by
factors that are Darboux polynomials!

For its importance and the fact that this result will be used
below, let us state it as a lemma:

\bigskip\bigskip\bigskip
\noindent
{\bf lemma:} \textit{If a polynomial $\Phi$ is written
as $\prod^{n_q}_j\phi_j^{k_j}$, where $\phi_j$ are irreducible
polynomials and $k_j$ integers, and
$\frac{D[{\Phi}]}{{\Phi}}=\Lambda$, where $\Lambda$ is a
polynomial and $D$ is a differential operator, with polynomial
coefficients, we have that:
$$ {\hbox{\scriptsize $\phi$}}_i | D[{\hbox{\scriptsize $\phi$}}_i], i=1 \cdots n_q$$ }

Let us now concentrate on the other part (the initial one) of
equation (\ref{44}) and, after some algebra, one gets:

\begin{eqnarray}
\label{50}
{\frac{D\left[\frac{{\cal B}}{\cal C}\right]}{ \left(
\frac{\cal B}{\cal C} \right) }} &=&\wp_1 (x,y,z) \Rightarrow \nonumber \\
D\left[{\cal B}\right] {\cal C} - D\left[{\cal C}\right] {\cal B}
&=& \wp_1 (x,y,z)\,{\cal C}\,{\cal B}\Rightarrow \nonumber \\
\frac{D\left[{\cal B}\right]}{{\cal B}} - \frac{D\left[{\cal
C}\right]}{{\cal C}} &=& \wp_1 (x,y,z)\Rightarrow \nonumber \\
\wp_1 (x,y,z)\,{\cal B} - D\left[{\cal B}\right] &=&-
\frac{D\left[{\cal C}\right]}{ {\cal C}}\,{\cal B}
\end{eqnarray}

Regarding the last line of equation (\ref{50}) above, we can see
that the left-hand side is obviously a polynomial. Since
$\frac{{\cal B}}{{\cal C}}$, by construction, is not a polynomial
(otherwise it would heve been simplified already) one can conclude
that:

\begin{equation}
\label{51} \frac{D[{\cal C}]}{{\cal C}} \,\,\,\,\hbox{is a
polynomial}
\end{equation}

Using this information and the third statement of equation
(\ref{50}) it is straightforward to conclude that

\begin{equation}
\label{52} \frac{D[{\cal B}]}{{\cal B}} \,\,\,\,\hbox{is also a
polynomial}
\end{equation}

So, using the above lemma, one may conclude that: If ${\cal B}$ is
written as  ${\cal B}= \prod^{n_b}_jb_j^{\kappa_j}$ and ${\cal} C$
as ${\cal C} = \prod^{n_c}_jc_j^{\lambda_j}$, where $b_j$ and
$c_j$ are irreducible polynomials and $\kappa_j$ and $\lambda$
integers, we may conclude that:

\begin{equation}
{\scriptsize b}_i | D[{\scriptsize b}_i], i=1 \cdots \kappa_q
\end{equation}

and

\begin{equation}
{\scriptsize c}_i | D[{\scriptsize c}_i], i=1 \cdots \lambda_q
\end{equation}

So, summarizing the analysis above,  for the case where $\left(
\prod_i p_i \right)$ does not have common factors with ${\cal
D}^2$ we have that ${\cal D}$, ${\cal B}$ and ${\cal C}$ are all
built from Darboux polynomials.

\subsubsection{$\left( \prod_i p_i \right)$ does have common
factors with ${\cal D}^2$}
\label{4.1.2}

In order to analyze this case, let us start by writing:

\begin{equation}
\label{53} \left( \prod_i p_i \right)=\left( \prod_{i=1}^{M} p_i
\right)= \left( \prod_{i=1}^{N} p_i \right).\left(
\prod_{i=N+1}^{M} p_i \right)=\eta . \zeta
\end{equation}
and

\begin{equation}
\label{54} {\cal D}=\left( \prod_{i=1}^{N'} d_i \right).\left(
\prod_{i=1}^{N} p_i \right)=\theta . \eta
\end{equation}
where

\begin{eqnarray}
\label{55}
\eta&=&\left( \prod_{i=1}^{N} p_i \right) \nonumber \\
\zeta&=&\left( \prod_{i=N+1}^{M} p_i \right) \nonumber \\
\theta&=&\left( \prod_{i=1}^{N'} d_i \right)
\end{eqnarray}

Now, from equations (\ref{42},\ref{53},\ref{54},\ref{55}), we can
write:
\begin{eqnarray}
\label{56} \left( \prod_i p_i \right) \left[\frac{ D[{\cal
A}]{\cal D} - {\cal A} D[{\cal D}] }{{\cal D}^2} \right] &=& \eta
. \zeta \left[\frac{ D[{\cal A}]{\cal D} - {\cal A} D[{\cal D}]
}{{\cal
D}\theta \eta} \right] \Rightarrow \nonumber \\
\zeta \left[\frac{ D[{\cal A}]{\cal D} - {\cal A} D[{\cal D}]
}{{\cal D}\theta} \right]&=& \wp (x,y,z)
\end{eqnarray}
where $\wp (x,y,z)$ is a polynomial.

After a little algebra, equation(\ref{56}) leads to:
\begin{eqnarray}
\label{57} \zeta D[{\cal A}]-\zeta {\cal A} \frac{D[{\cal
D}]}{{\cal D}}&=&\wp (x,y,z) \theta \Rightarrow \nonumber \\
\zeta {\cal A} \frac{D[{\cal D}]}{{\cal D}}&=& \verb"polynomial"
\end{eqnarray}

Since, by construction, ${\cal A}$ does not have any factors in
common with ${\cal D}$ and, analogously, $\zeta$ is the part of
$\left( \prod_i p_i \right)$ that ``is not'' on ${\cal D}$, one
may conclude that:

\begin{equation}
\label{58}
\frac{D[{\cal D}]}{{\cal D}}=\flat(x,y,z) =
\verb"polynomial"
\end{equation}

So, again, by using the lemma proved above, If we write ${\cal D}$
as $\prod^{n_q}_jd_j^{k_j}$, where $d_j$ are irreducible
polynomials and $k_j$ integers, we have:
$$ {\hbox{\scriptsize
d}}_i | D[{\hbox{\scriptsize d}}_i], i=1 \cdots n_q$$ }.

\noindent Therefore, as in the case of the previous subsection,
${\cal D}$ can be written as a product of Darboux polynomials.

\bigskip\bigskip
Let us now use equations (\ref{dinv_expandido
2},\ref{dinv_expandido dois}). Using these, one can write:

\begin{eqnarray}
\label{59} 
{\frac{D[\prod_i p_i^{c_i}]}{ \left( \prod_i p_i^{c_i}
\right) }} + D\left[\frac{{\cal A}}{\cal D }\right]&=&0
\Rightarrow
\nonumber \\
{\frac{\sum_j c_j D[p_j] \left(\prod_{k \neq j}
p_k\right)}{\prod_i p_i}} + D\left[\frac{{\cal A}}{\cal D }\right]
&=&0 \Rightarrow
\nonumber \\
\sum_j {\frac{c_j D[p_j]}{p_j}} + D\left[\frac{{\cal A}}{\cal D
}\right] &=& 0
\end{eqnarray}

Now, remembering that, we are considering the case where ${\cal
D}$ has factors in common with $\left( \prod_i p_i \right)$
(equations (\ref{53},\ref{54})) and that we have just proven that
${\cal D}$ can be written as a product of Darboux polynomials, we
can write:

\begin{eqnarray}
\label{60}
\underbrace{\sum_{j=1}^{N} {\frac{c_j D[p_j]}{p_j}}} +
\sum_{j=N+1}^{M} {\frac{c_j D[p_j]}{p_j}} + D\left[\frac{{\cal
A}}{\cal D }\right] &=& 0 \Rightarrow \nonumber \\
\wp_1 \verb"(polynomial)"
\,\,\,\,\,\,\,\,\,\,\,\,\,\,\,\,\,\,\,\,\,\,\,\,\,\,\,\,\,\,\,\,\,\,\,\,\,\,\,\,\,\,\,\,\,\,\,\,\,\,\,\,\,\,\,\,\,\,\,\,\,\,\,
&& \nonumber \\
&& \nonumber \\
D\left[\frac{{\cal A}}{\cal D }\right] + \wp_1 = -
{\frac{\sum_{j=N+1}^M c_j D[p_j] \left(\prod_{k \neq j}
p_k\right)}{\prod_{i=N+1}^M p_i}} &&
\end{eqnarray}
\noindent where the last line was obtained by re-writing the
summation with a single denominator.

Finally, one can conclude that:
\begin{eqnarray}
\label{61} D\left[\frac{{\cal A}}{\cal D }\right] + \wp_1 = -
{\frac{\sum_{j=N+1}^M c_j D[p_j] \left(\prod_{k \neq j}
p_k\right)}{\prod_{i=N+1}^M p_i}}&\Rightarrow & \nonumber \\
D\left[\frac{{\cal A}}{\cal D }\right]\left( \prod_{i=N+1}^M
p_i\right) = - \wp_1 \left( \prod_{i=N+1}^M p_i\right)-
\sum_{j=N+1}^M c_j D[p_j] \left(\prod_{k \neq j} p_k\right)
\end{eqnarray}
A direct inspection of the equation above allows for the
conclusion that:
\begin{equation}
D\left[\frac{{\cal A}}{\cal D }\right]\left( \prod_{i=N+1}^M
p_i\right) = \wp_2 = \verb"polynomial"
\end{equation}
Expanding the left-hand side of the equation above and considering
that, by construction, $\left( \prod_{i=N+1}^M p_i\right)$ does
not have any common factor with ${\cal D}$, we have
\begin{eqnarray}
\label{64}
\left[\frac{ D[{\cal A}]{\cal D} - {\cal A} D[{\cal D}]
}{{\cal D}^2} \right] \left( \prod_{i=N+1}^M p_i\right) =
\wp_2\Rightarrow
&& \nonumber \\
D\left[\frac{{\cal A}}{\cal D }\right] = \wp_3 = \verb"polynomial"
&&
\end{eqnarray}
So, using this, and the reasoning following equation (\ref{44}),
we may conclude that, for the situation being studied here (i.e.,
$\left( \prod_i p_i \right)$ does have common factors with ${\cal
D}^2$ ), we also have that
\begin{eqnarray}
\label{50-b} {\frac{D\left[\frac{{\cal B}}{\cal C}\right]}{
\left( \frac{\cal B}{\cal C} \right) }}
=\emptyset(x,y,z)&=&\verb"polynomial" \Rightarrow \nonumber \\
\frac{D[{\cal B}]}{{\cal B}} \,\,\,\,\hbox{is a
polynomial}&\verb"and"&\frac{D[{\cal C}]}{{\cal C}}
\,\,\,\,\hbox{is a polynomial}
\end{eqnarray}
So, by the lemma proved above and by equations
(\ref{DDarboux},\ref{50-b}), we may infer that:

If we write ${\cal D}$ as $\prod^{n_q}_jd_j^{k_j}$, ${\cal B}$ as
$\prod^{m_q}_jb_j^{\kappa_j}$ and ${\cal C}$ as
$\prod^{\mu_q}_jc_j^{\nu_j}$, where $d_j$, $b_j$ and $c_j$ are
irreducible polynomials and $k_j$, $\kappa_j$ and $\nu_j$ integers
we have:

\begin{equation}
\label{66} {\scriptsize b}_i | D[{\scriptsize b}_i], i=1 \cdots
n_q\verb",  " {\scriptsize b}_i | D[{\scriptsize b}_i], i=1 \cdots
m_q\verb"  and "{\scriptsize b}_i | D[{\scriptsize b}_i], i=1
\cdots \mu_q
\end{equation}

So, also for the case where $\left( \prod_i p_i \right)$ does have
common factors with ${\cal D}^2$ we have that ${\cal D}$, ${\cal
B}$ and ${\cal C}$ are built from Darboux polynomials.

\bigskip
\noindent
So, summing up the results for sections
(\ref{4.1.1},\ref{4.1.2}), we conclude that for a differential
invariant given by equation (\ref{Inv form 2}), the following
result is demonstrated:

\bigskip\bigskip\bigskip
\noindent
{\bf Theorem:} \textit{For a differential invariant of
the form ${\cal I} = e^{\left( \frac{A}{{\cal D}} \right)}\left( \frac{B}{C} \right)$, where $A, {\cal D}$, ${\cal B}$ and ${\cal C}$ are all formed by products of independent, irreducible
polynomials (i.e., the fractions can not be simplified further)
such that ${\cal D}$ as $\prod^{n_q}_jd_j^{k_j}$, where $d_j$ are
irreducible polynomials and $k_j$ integers,  ${\cal B}$ as
$\prod^{m_q}_jb_j^{\natural_j}$, where $b_j$ are irreducible
polynomials and $\natural_j$ integers and  ${\cal C}$ as
$\prod^{l_q}_jc_j^{\flat_j}$, where $c_j$ are irreducible
polynomials and $\flat_j$ integers, we have that the $d_j, b_j$
and $c_j$ are all Darboux polynomials of the operator $ D \equiv N\,\partial_{x} + z\,N \partial_{y} + M\, \partial_{z} $}

\bigskip
So, combining all the above discussion, equation (\ref{Inv_form})
leads to:
\begin{equation}
\label{Inv_form_ABCD} \phi = \frac{\left( B_{{x}}+zB_{{y}} \right)
C{{\cal D}}^{2}+ \left( -C_{{x}}-zC_{{y}} \right) B{{\cal D}}^{2}+
\left( A_{{x}}+zA_{{y}} \right) BC{\cal D}+ \left( -A{\cal
D}_{{x}}-zA{\cal D}_{{y}} \right) BC }{-A_{{z}}{\cal D}\,CB+A{\cal
D}_{{z}}CB-{{\cal D}}^{2}B_{{z}}C+{{\cal D}}^{2}BC_{{z}}}
\end{equation}
\bigskip

As we shall see in the next sections of the paper, the information
contained on (\ref{Inv_form_ABCD}) will allow for finding
``hidden'' Darboux polynomials.

\subsection{Darboux polynomials as functions of $(z)$, present on the denominator}
\label{Nz}

One might ask what is the point of the title for this subsection.
As we have shown previously, from the denominator of the ODE (or
from its factors), if it (they) is (are) function(s) of $(x,y)$
only, we can extract Darboux polynomials straightforwardly. We have shown in section
(\ref{Nxy}) that this information might prove crucial to reducing
the ODE in question. So the question that one may ask is if it is
possible to extract information regarding Darboux polynomials
depending on $z$ from the denominator of the ODE. Using (\ref{Inv_form_ABCD}) one can.

Let us consider the denominator of the ODE as written on
(\ref{Inv_form_ABCD}):

\begin{equation}
\label{denom_INV}
N = -A_{{z}}{\cal D}\,CB+A{\cal
D}_{{z}}CB-{{\cal D}}^{2}B_{{z}}C+{{\cal D}}^{2}BC_{{z}}
\end{equation}

Of course, that is, in principle, the general expression for the
denominator of the ODE. Surely, it can happen that, for some
particular combination of $A,{\cal D}, B$ and $C$, some factor of
(\ref{denom_INV}) will be also a factor on the numerator for the
ODE and some cancellation will occur. But, disregarding this
possibility for now, let us particularize the analysis a little.

Let us consider the particular situation where:
\begin{eqnarray}
\label{case_one} {\cal D}&=&k (constant) \nonumber \\
A&=&A(x,y)\rightarrow A_z=0
\end{eqnarray}
For this case, equation (\ref{denom_INV}) becomes:
\begin{equation}
\label{denom_INV_reduced}
N = -{{\cal D}}^{2}B_{{z}}C+{{\cal
D}}^{2}BC_{{z}} \rightarrow k^{2}(-B_{z}C+BC_{z})
\end{equation}
\noindent Consider the case where:
\begin{eqnarray}
\label{case_one_BzCz}
B(x,y,z)&=& {\cal B}(z)+b(x,y) \nonumber \\
C(x,y,z)&=& {\cal C}(z)+c(x,y)
\end{eqnarray}
where ${\cal B},{\cal C},b$ and $c$ are polynomials.

So, combining these results (eqs. (\ref{case_one}) and
(\ref{case_one_BzCz})), we may finally conclude that:
\begin{eqnarray}
\label{N_INV_reduced} N &= &k^{2}(-B_{z}C+BC_{z})\rightarrow
-k^2\left( {\frac {d}{dz}}B \left( z \right)  \right) C \left( z
\right) -k^2 \left( {\frac {d}{dz}}B \left( z \right)  \right) c
\left( x,y
 \right) + \nonumber \\
 && + k^2\left( {\frac {d}{dz}}C \left( z \right)  \right) B \left(
z \right) + k^2\left( {\frac {d}{dz}}C \left( z \right)  \right) b
 \left( x,y \right)
\end{eqnarray}

It seems that the situation has not improved. But, take heart, if
we further consider that
\begin{equation}
\label{BzCz}
B(z)=C(z)={\cal K}(z),
\end{equation}
 we finally find that:
\begin{eqnarray}
\label{N_INV_reduced_dois} N &= & -k^2\left( {\frac {d}{dz}}B
\left( z \right)  \right) c \left( x,y \right)  +k^2 \left( {\frac
{d}{dz}}C \left( z \right) \right) b
 \left( x,y \right)\rightarrow \nonumber \\
 && \rightarrow k^2\left( {\frac {d}{dz}}{\cal K} \left( z
\right)  \right) \left( b(x,y)-c(x,y) \right)
 \end{eqnarray}

Actually, a little variation of this idea could broaden the range
for the application of it. Consider that, instead of equation
(\ref{BzCz}) we had:
\begin{equation}
\label{BzCz_modified} c_B\,B(z)=c_C\,C(z)=c_B\,c_C\,{\cal K}(z),
\end{equation}
where $c_B$ and $c_C$ are constants. In turn, that will imply that

\begin{eqnarray}
\label{N_INV_reduced_tres}
B_z&=&c_C\,{\cal K}_z \nonumber \\
C_z&=&c_B\,{\cal K}_z \rightarrow \nonumber \\
\rightarrow N&=&  k^2\left( {\frac {d}{dz}}{\cal K} \left( z
\right) \right) \left( c_C\,b(x,y)-c_B\,c(x,y) \right)
\end{eqnarray}
If we use $c_B=c_C=$ in (\ref{BzCz_modified}) we recover
(\ref{BzCz}). Thus, using (\ref{BzCz_modified}) generalizes
(\ref{BzCz}). This can be very useful, as we shall see on section
(\ref{examples_section_4}).

So, the main trust of the approach suggested in
 this section can be summarize as an algorithm in the following steps:

\bigskip
\bigskip
\centerline{\bf Steps of the Algorithm}
  \begin{enumerate}
 \item Inspecting the denominator of the 2ODE,
 analyzing equation (\ref{N_INV_reduced_dois}), determine candidates for the functions ${\cal K}(z)$, $B(x,y,z)$ and $C(x,y,z)$ (eq.
 (\ref{case_one_BzCz})).
 \item Considering that the conditions of sub-section \ref{4ponto1} are met, these functions would be then, each, Darboux polynomials. Verify that.
 \item If the above verification proves true, run the algorithm introduced in \cite{JMP}.
 \end{enumerate}

\noindent
\bigskip
 An example of the application of these ideas will be presented in
 section (\ref{examples_section_4}).

\subsection{Darboux polynomials as functions of $(x,y)$, present on the numerator}
\label{Mxy}

Here we are going to do something similar to the procedure shown
above (where we have extracted information enabling us to
determine Darboux polynomials depending on $z$ from the
denominator of the corresponding ODE). We are going to study the
possibility of extracting Darboux polynomials (deppending on $(x,y)$) from
the numerator of the ODE.

Let us analyze equation (\ref{Inv_form_ABCD}). Restricting
ourselves to the case:

\begin{eqnarray}
\label{case_two}
{\cal D} &=& k\, (constant) \nonumber \\
A &=& A(z) \rightarrow \,\, A_x=A_y=0,
\end{eqnarray}
the numerator for the ODE (see (\ref{Inv_form_ABCD}) becomes:

\begin{equation}
\label{numer_INV}
M = \left( B_{{x}}+zB_{{y}} \right) C{{\it
k}}^{2}+ \left( -C_{{x}}-zC_{{y}} \right) B{{\it k}}^{2}
\end{equation}

Surely, it can
happen that, for some particular combination of $A,{\cal D}, B$
and $C$, some factor of (\ref{numer_INV}) will be also a factor on
the denominator for the ODE and some cancellation will occur. But,
disregarding this possibility for now, let us particularize the
analysis a little further.

Consider the following:
\begin{eqnarray}
\label{case_two_BxyCxy}
B(x,y,z)&=& {\cal \beta}(x,y)+\flat(z) \nonumber \\
C(x,y,z)&=& {\cal \gamma}(x,y)+\natural(z)
\end{eqnarray}
where ${\cal \beta},{\cal \gamma},\flat$ and $\natural$ are
polynomials.

Following the lead from sub-section (\ref{Nz}), equation
(\ref{BzCz}), considering:
\begin{equation}
\label{BxyCxy}
{\cal \beta}(x,y)={\cal \gamma}(x,y)=\theta (x,y),
\end{equation}
after a little algebra, from equation (\ref{numer_INV}) one gets:
\begin{eqnarray}
\label{numer_INV_reduced}
M&=&\left( B_{{x}}+zB_{{y}} \right)
C{{\it k}}^{2}+ \left(
-C_{{x}}-zC_{{y}} \right) B{{\it k}}^{2} \rightarrow \nonumber \\
&\rightarrow&{{\it k}}^{2} \left(\theta_x+z\,\theta_y\right)
\left(\natural(z) -\flat(z)\right)
\end{eqnarray}

Again, we may summarize the procedure here introduced to try and find Darboux polynomials, as functions of $(x,y)$, from the numerator of the ODE under consideration, as an simple algorithm:

\bigskip
\bigskip
\centerline{\bf Steps of the Algorithm}
\begin{enumerate}
 \item Inspecting the numerator of the 2ODE,
 analyzing equation (\ref{numer_INV_reduced}), determine candidates for the functions $\theta(x,y)$, $\natural (z)$ and $\flat (z)$.
 \item Considering that the conditions of sub-section \ref{4ponto1} are met, these functions would be then, each, Darboux polynomials. Verify that.
 \item If the above verification proves true, run the algorithm introduced in \cite{JMP}.
\end{enumerate}

In the next section (\ref{examples_section_4}), we will present an
example where this equation is put into use and provides the
finding of Darboux polynomials.

\subsection{Examples}
\label{examples_section_4}

Although, as mentioned above. it is not the point of this paper to
compare the capabilities of the method presented on \cite{JMP}
with other methods and algorithms that deal with ODEs, it is
interesting to mention, in the sense of displaying the practical
usage of the ideas here presented, that both examples on this
section can not be solved by the powerful methods implemented on
the Maple symbolic basin (release 10).

\subsubsection{third example}
Here we are going to exemplify the usefulness of the ideas and
equations developed in section (\ref{Nz}).

Consider the following ODE:

\begin{equation}
\label{thirdODE} {\frac {d^{2}y}{d{x}^{2}}} = {\frac
{-c{z}^{10}a-6\,{z}^{6}d{y}^{5}a+ \left( -bc+ad{y}^{6}-cb x
\right) {z}^{5}-6\,zd{y}^{5}bx+d{y}^{6}bx+bd{y}^{6}}{(-5{z}^{4})
\left( ad{y}^{6}+cbx \right) }}
\end{equation}
Comparing this above equation with equation
(\ref{N_INV_reduced_dois}), we can try and deduce two Darboux
polynomials from ${z}^{4} \left( ad{y}^{6}+cbx \right)$.

The most obvious ``guess'' is that, for example:
\begin{eqnarray}
k&=&1 \nonumber \\
{\cal K}(z)&=&z^5 \nonumber \\
b(x,y)&=&a\,d\,y^6 \nonumber \\
c(x,y)&=&-c\,b\,x
\end{eqnarray}

Unfortunately, this does not lead to actual Darboux polynomials.
So it seems that the analysis presented on section (\ref{Nz}) does
not help us here. But, as previously advertised, if we use
equation (\ref{BzCz_modified}), we would be lead to make a
different ``guess'' and that might result successful. Trying
\begin{eqnarray}
k&=&1 \nonumber \\
{\cal K}(z)&=&z^5 \nonumber \\
B(z)&=&a\,{\cal K}(z) \nonumber \\
C(z)&=&c\,{\cal K}(z) \nonumber \\
b(x,y)&=&d\,y^6 \nonumber \\
c(x,y)&=&-b\,x
\end{eqnarray}
we find a couple of Darbox polynomials. We present them and the
corresponding co-factors bellow:
\begin{eqnarray}
v_1=-c{z}^{5}+d{y}^{6}& &g_1=5\,
{z}^{4} \left( 6\,zd{y}^{5}a+{z} ^{5}ac+cbx+bc \right) \nonumber \\
v_2={z}^{5}a+bx&&g_2=-5\, \left(
-6\,zd{y}^{5}a-bc+ad{y}^{6}-{z}^{5}ac \right) {z}^{4}
\end{eqnarray}

As usual, using these results, running the methods and algorithms
introduced on \cite{JMP}, we can find:
\begin{eqnarray}
\label{RPQTRES}
P&=& -6\,d{y}^{5} \left( {z}^{5}a+bx \right)\nonumber \\
Q&=& 1 \nonumber \\
R&=&\frac{1}{\left( -c{z}^{5}+d{y}^{6} \right)  \left(
{z}^{5}a+bx \right)}.
\end{eqnarray}
and, using (\ref{IxIyIz2}) and (\ref{IntI}), we finally find:
\begin{equation}
\label{InvTRES}
 I = -x+\ln  \left( -c{z}^{5}+d{y}^{6} \right) -\ln  \left( a{z}^{5}+bx
 \right).
\end{equation}

From (\ref{RPQTRES}), it is easy to see that we needed the two
Darboux polynomials we have found using the technique displayed on
section (\ref{Nz}) in the sense that the integrating factor $R$ is
constructed from both of them.

\subsubsection{fourth example}
In this section, we are going to exemplify the use of the results
listed on section (\ref{Mxy}). Let us get to work by examining the
following ODE:

\begin{equation}
\label{ode4example} {\frac {d^{2}y}{d{x}^{2}}} = {\frac {- \left(
a{z}^{2}-z-2\,az+a \right)  \left( 4\,z{y}^{3}+3\,b{x }^{2}
\right) }{ \left( b{x}^{3}+{y}^{4} \right)  \left( 3\,{z}^{4}a-3
\,{z}^{3}-6\,{z}^{3}a+3\,{z}^{2}b{x}^{3}+3\,{z}^{2}{y}^{4}+2\,a{z}^{2}
+z+a-b{x}^{3}-{y}^{4}+1 \right) }}
\end{equation}

\bigskip
Using the techniques suggested on sections (\ref{Nxy}) and
(\ref{Mz}), we can deduce that $\left( b{x}^{3}+{y}^{4} \right)$
and $\left( a{z}^{2}-z-2\,az+a \right)$ are Darboux polynomials of
the $D$ operator (equation (\ref{Doperator})) corresponding to the
ODE (\ref{ode4example}). Indeed, this is true and very quick. The
question is: is that enough to find an integrating factor (and,
consequently, a differential invariant) for (\ref{ode4example}),
using the Darboux method we are using here \cite{JMP} The answer
is no. So, let us ask section (\ref{Mxy}) for help.

Basically, we want to compare equation (\ref{numer_INV_reduced})
with the numerator of the ODE (\ref{ode4example}).

\begin{eqnarray}
\label{application4}
{{\it k}}^{2}
\left(\theta_x+z\,\theta_y\right) \left(\natural(z)
-\flat(z)\right) &\Leftrightarrow&\left( a{z}^{2}-z-2\,az+a
\right)  \left( 4\,z{y}^{3}+3\,b{x }^{2} \right) \nonumber \\
&&
\end{eqnarray}

Surely, there are a lot of possible sets of $\flat(z),
\natural(z), k$ and $\theta(x,y)$ to satisfy equation
(\ref{application4}). A satisfactory answer (one producing Darboux
polynomials in enough numbers to find a differential invariant for
the ODE under study) is found for:

\begin{eqnarray}
\label{goodset}
k&=&0 \nonumber \\
\flat(z)&=&0 \nonumber \\
\natural(z)&=& \left( a{z}^{2}-z-2\,az+a \right) \nonumber \\
\theta(x,y)&=&\left( b{x}^{3}+{y}^{4} \right)
\end{eqnarray}

With this solution to (\ref{application4}), we find the following
two Darboux polynomials and corresponding co-factors:
\begin{equation}
\begin{tabular}{lll}
  $v_1= \left(b{x}^{3}+{y}^{4}\right) \rightarrow $& &  \\
   $g_1=\left(
3\,b{x}^{2}+4\,z{y}^{3} \right) \left(
3\,{z}^{4}a-6\,{z}^{3}a-3\,{z}^{3}+3\,{z}^{2}b{x}^{3}
+3\,{z}^{2}{y}^{4}+2\,a{z}^{2}+z+a-b{x}^{3}-{y}^{4}+1 \right)$&& \\
\\
$v_2=\left(b{x}^{3}+{y}^{4}+a{z}^{2}-2\,az+a-z\right) \rightarrow$ & &  \\
$ g_2= -\left( 3\,b{x}^{2}+4\,z{y}^{3} \right) \left( b{x}^{3}-3\,{z}^{2}b{x}^{3}-1+{y}^{4}-2\,a+2\,az-3\,{z}^{2}{y}^{4} \right)$ &  &
\end{tabular}
\end{equation}
using the method briefly explained in section (\ref{intro}), we
calculate

\begin{eqnarray}
\label{RPQQUATRO}
P&=&4\,{y}^{3} \left( a{z}^{2}-z-2\,az+a \right) \nonumber \\
Q&=& 1 \nonumber \\
R&=&{\frac {1}{ \left( b{x}^{3}+{y}^{4} \right)  \left(
b{x}^{3}+{y}^{4}+a{z}^{2}-2 \,az+a-z \right) }}  .
\end{eqnarray}
and, finally:
\begin{equation}
\label{InvQUATRO}
 I = {z}^{3}-z+\ln  \left( {\frac {b{x}^{3}+{y}^{4}}{b{x}^{3}+{y}^{4}+a{z}^
{2}-2\,az+a-z}} \right).
\end{equation}

\section{Conclusion}
\label{conclusion}

From our experience working on the Darboux type methods to deal
with ODEs, we concluded that,  on all the methods and algorithms,
the necessary step of finding the Darboux polynomials is always
very costly computationally. So, we have developed methods to
extract those polynomials from the ODEs themselves.

Some are very straightforward once we analyze the differential
operator $D \equiv N\,\partial_{x} + z\,N \partial_{y} + M\,
\partial_{z},$ used on the procedure we have presented on
\cite{JMP} and, here, used as our example os methods benefitting
from a quick finding of Darboux polynomials. Actually, as we have
mentioned on the paper, this feature of the procedure developed on
\cite{JMP} is very useful. If $N$ is a function of $(x,y)$ it will
be itself a Darboux polynomial and the same applies for any factor
of $N$ that is a polynomial on $(x,y)$. Analogous results apply
for $M$ (and its factors) that are polynomials on $(z)$. Even this
simple analysis allows that, in practice, one can use a Darboux
type approach where a high degree Darboux polynomial is needed
(say third degree and up) in a practical and feasible way.

On the last sections of the paper, we have presented other
examples of the idea of trying to extract the Darboux polynomials
inspecting the ``format'' of the ODE. This time, a more elaborate
analysis had to be performed. We have used the fact that
the invariant is of the form:
\begin{eqnarray}
\label{Inv_form_conclusion} I &= &{\frac {A \left( x,y,z \right)
}{{\cal D} \left( x,y,z \right) }}+\ln
 \left( {\frac {B \left( x,y,z \right) }{C \left( x,y,z \right) }}
 \right)
\end{eqnarray}
where $A \left( x,y,z \right),{\cal D} \left( x,y,z \right) , B
\left( x,y,z \right)$ and $C \left( x,y,z \right)$ are all
polynomials and, furthermore, ${\cal D} \left( x,y,z \right), B
\left( x,y,z \right)$ and $C \left( x,y,z \right)$ are Darboux
polynomials of the D-operator just mentioned. Of course, that is not the most general case (as previously stated). But, from our experience, this is not a very restricted case (in comparison to the stringent demand we have already imposed of considering only rational ODEs). From this we have the general expression for the ODE and,
analyzing some classes of equations, were able to determine the
needed Darboux polynomials for examples belonging to the class.

With this, we hope to have establish that this kind of approach is
very important to the understanding of the structure of ODEs and
to make the solving (reducing) of them, via Darboux methods, more
practical in cases where high degree polynomials are present.

Many roads are open to follow these ideas. We are currently
pursuing some of them. For instance, we are working on a
classificatory system for 2ODEs based on equation
(\ref{Inv_form_ABCD}) and the ideas here exposed.

\newpage
%%%aqui

%%%%%%%%%%%%%%%%%%%%%%%%%%%%%%

\end{document}